\documentclass[sigconf]{acmart}

\usepackage{enumitem}
\usepackage{cleveref}
\usepackage{multirow}

\copyrightyear{2026}
\acmYear{2026}
\setcopyright{cc}
\setcctype{by-nc-nd}
\acmConference[GLSVLSI '26]{Great Lakes Symposium on VLSI 2026}{June 22--24, 2026}{Canandaigua, NY, USA}
\acmBooktitle{Great Lakes Symposium on VLSI 2026 (GLSVLSI '26), June 22--24, 2026, Canandaigua, NY, USA}
\acmDOI{10.1145/3787109.3815288}
\acmISBN{979-8-4007-2431-2/2026/06}

\begin{document}

\title{DORA: \underline{D}ataflow-Instruction \underline{Or}chestration \underline{A}rchitecture \\ for DNN Acceleration}

\settopmatter{authorsperrow=4}
\author{Xingzhen Chen}
\affiliation{%
  \institution{Brown University} 
  \city{Providence}
  \country{USA}
  }
\email{xingzhen_chen@brown.edu}

\author{Zhuoping Yang}
\affiliation{%
  \institution{Brown University} 
  \city{Providence}
  \country{USA}
  }
\email{zhuoping_yang@brown.edu}

\author{Jinming Zhuang}
\affiliation{%
  \institution{Brown University} 
  \city{Providence}
  \country{USA}
  }
\email{jinming_zhuang@brown.edu}

\author{Shixin Ji}
\affiliation{%
  \institution{Brown University} 
  \city{Providence}
  \country{USA}
  }
\email{shixin_ji@brown.edu}

\author{Sarah Schultz}
\affiliation{%
  \institution{Brown University} 
  \city{Providence}
  \country{USA}
  }
\email{sarah_schultz2@brown.edu}

\author{Zheng Dong}
\affiliation{%
  \institution{Wayne State University}
  \city{Detroit}
  \country{USA}
  }
\email{dong@wayne.edu}

\author{Weisong Shi}
\affiliation{%
  \institution{University of Delaware}  
  \city{Newark}
  \country{USA}
  }
\email{weisong@udel.edu}

\author{Peipei Zhou}
\affiliation{%
  \institution{Brown University} 
  \city{Providence}
  \country{USA}
  }
\email{peipei_zhou@brown.edu}

\renewcommand{\shortauthors}{
Xingzhen Chen,
Zhuoping Yang,
Jinming Zhuang,
Shixin Ji,\\
Sarah Schultz,
Zheng Dong,
Weisong Shi,
and Peipei Zhou
}


\begin{abstract}

As deep neural networks develop significantly more diverse and complex, achieving high performance and efficiency on complicated DNN models faces pressing challenges. Modern DNN workloads are increasingly diverse in operation types, tensor shapes, and execution dependencies, making it difficult to sustain high hardware efficiency across models. In addition, a generic accelerator often incurs substantial overhead when executing diverse workloads.

To address these problems, we propose DORA, an instruction-based overlay architecture that explicitly describes dataflow via a proposed ISA, enabling fine-grained control of data movement, computation, and synchronization at the layer level.
To support flexibility while achieving high performance, DORA adopts a novel on-chip memory management and computation parallelism management mechanism. 
DORA proposes a compilation framework that can generate instructions for given DNN workloads after a two-stage design space exploration. 
DORA framework also incorporates a MILP-based and a heuristic-based search engine to generate the schedule solution for different needs and constraints.

We prototype DORA on the AMD Versal VCK190 platform, demonstrating its deployability on existing reconfigurable systems. Experimental results show that DORA maintains stable efficiency, with less than 5\% variation on a single vector processor across workloads exhibiting up to 6$\times$ variation in operation counts. Compared to state-of-the-art accelerators, DORA consistently achieves higher performance, delivering up to 5$\times$ throughput improvement. The heuristic-based scheduler further achieves up to 90\% optimality under practical time constraints. DORA is open-sourced at \url{https://github.com/arc-research-lab/DORA.git}.

\end{abstract}


\maketitle

\section{Introduction}

With the development of deep neural networks (DNNs), the end-to-end perception tasks in autonomous driving systems (ADS) \cite{autoware} increasingly rely on diverse DNN models for image segmentation \cite{DeiT}, classification \cite{MLPMixer}, and 3-D point-cloud processing \cite{qi2017pointnet}, which vary a lot in model structure and model size.
GPUs \cite{cuda_programming} are general accelerators with black-box scheduling, which fail in deterministic scenarios \cite{ji2025art}.
To meet the stringent performance and efficiency requirements of modern DNN inference, hardware accelerators are widely adopted. There are two trends of accelerator design methodologies: dedicated accelerators~\cite{zhuang2023charm,ssr,chen2024understanding,hall2020hpipe,liu2025flightvgm,zeng2024flightllm,dong2024eq,zhang2020dnnexplorer,kwon2021herald,cai2023set,PEGASO,Menzel} and overlay-based accelerators~\cite{RSN,tong2024feather,yang2025nsflow,he2025intar,abdelfattah2018dla,zhang2022fast,BeyondPeakPerformance,Cloud-Scale,guo2024overlay,wei2018tgpa}.
The dedicated accelerators rely on static dataflow and fixed-function calls, which often incur substantial overhead when handling diverse workloads. Dynamic reconfiguration on modern FPGAs, such as PCAP \cite{PCAP_partial_reconfig} and VERSATILE \cite{ibrahim2025versatile}, typically takes several to tens of milliseconds. Such overhead is unacceptable for autonomous driving applications that require processing frequencies of 30$\sim$100 Hz \cite{autoware,industrychallenge}, corresponding to a strict latency budget of only 10$\sim$33 ms per task.
In contrast, overlay-based architectures improve programmability and flexibility, but existing works still struggle to efficiently support highly diverse workloads.

\begin{figure}
    \centering
    \includegraphics[width=0.9\linewidth]{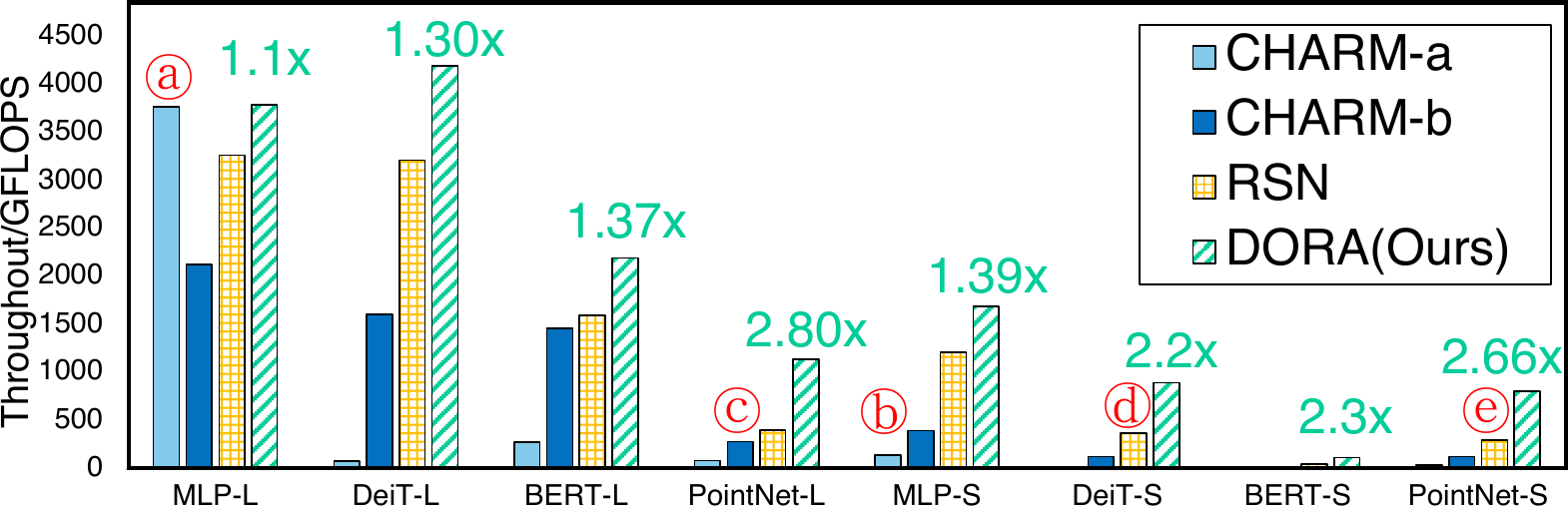}
    \caption{Profiling results for MLP, DeiT, BERT, and PointNet applications on CHARM 2.0, RSN, and DORA architectures. -L stands for large model, and -S stands for small model.}
    \label{fig:introduction}
\end{figure}

\setlength{\textfloatsep}{6pt}
\setlength{\floatsep}{5pt}
\setlength{\intextsep}{5pt}

Figure~\ref{fig:introduction} shows the profiling results of representative DNN models with large and small versions, on state-of-the-art dedicated accelerator, CHARM 2.0~\cite{charm2}, and overlay-based accelerator, RSN~\cite{RSN}.
MLP serves as a low-variance workload, where most layers are dominated by near-square matrix multiplications (MM) with relatively consistent dimensions. In contrast, DeiT and PointNet have much higher variance due to the presence of both large and small MM.
CHARM-a is a monolithic architecture generated from CHARM framework for all six models.
However, as illustrated at Point~\textcircled{a}, the layer shapes within MLP-L are not uniform, preventing CHARM-a from fully exploiting the available hardware resources to the same extent as DORA. Moreover, when layer dimensions are smaller than the fixed design size (e.g., the loop boundaries within vector processor are fixed), CHARM-a must perform off-chip padding, which introduces significant communication overhead and leads to under-utilization of computation parallelism, resulting in substantial performance degradation at Point~\textcircled{b}.
CHARM 2.0 also supports partitioning resources into two independent accelerators for diverse operand shapes, and we leverage it to generate an optimized solution, i.e., CHARM-b, for these models.
However, since the resource is statically partitioned at compilation time with a coarse-grained granularity, the performance still degrades significantly when the model has high variance (Point \textcircled{c}).
RSN introduces the instruction-based control method, which can flexibly switch on-chip dataflow by decoding instructions at runtime. However, RSN only explores the flexible mapping strategy for on-chip storage, and keeps the parallelism and buffering strategies tailored for specific-sized models (Point \textcircled{d}). Therefore, the performance drops rapidly when the layer shape becomes more diverse, and the model size becomes smaller than the designed buffer granularity (Point \textcircled{e}).

To address this problem, it is essential to enable layer-wise dataflow switching and flexible resource management with fine-grained, layer-level workload awareness, and fully utilize programming flexibility that is provided by vendor manuals \cite{amd_aie_intrinsics_ug}.
Moreover, building an end-to-end compilation tool for the instruction set architecture (ISA) is also important and challenging. Therefore, we propose DORA, a dataflow-instruction orchestration architecture, consisting of an ISA design and a compilation framework for diverse DNN models.
Our contributions are highlighted.


\vspace{-5pt}
\begin{itemize}[leftmargin=*]
    \item We propose an overlay architecture, featuring a variety of specialized functional units, which can be programmable via instructions during runtime to exploit the flexibility of the underlying hardware of memory and parallelism management on diverse DNN workloads, which is the first work to fully utilize the intrinsic flexibility provided by AI Engine.
    \item We propose a fully automated two-stage compilation framework that includes: (1) Dynamic resource allocation at the single-layer level; (2) Scheduling multiple layers to functional units at the graph level; and (3) Instruction generation for input workloads.
    \item DORA provides two options to accelerate the design space exploration (DSE), i.e., DAG Partition and Genetic Algorithm.
    \item Our prototype results demonstrate that DORA maintains stable efficiency for single vector processors, with less than 5\% variation across workloads exhibiting up to a 6$\times$ variation in operation counts. Compared to state-of-the-art accelerators, DORA consistently achieves higher performance, delivering up to a 5$\times$ throughput improvement. The heuristic-based scheduler further achieves up to 90\% optimality under practical time constraints.
    \item DORA is open-sourced at \url{https://github.com/arc-research-lab/DORA.git}.
    
\end{itemize}

\vspace{-7pt}
\section{Related Works}
\vspace{-2pt}

Overlay~\cite{RSN,tong2024feather,yang2025nsflow,he2025intar,abdelfattah2018dla,zhang2022fast,BeyondPeakPerformance,Cloud-Scale,guo2024overlay,wei2018tgpa} and static-dataflow accelerators~\cite{zhuang2023charm,ssr,chen2024understanding,hall2020hpipe,liu2025flightvgm,zeng2024flightllm,dong2024eq,zhang2020dnnexplorer,kwon2021herald,cai2023set} are two popular design styles in hardware accelerators for AI workloads.
RSN \cite{RSN} supports flexible allocation of functional units and optimizes for medium-sized models. InTAR \cite{he2025intar} adopts a fixed datapath, and the functionality of each PE keeps statically.
Bouaziz et al. \cite{MonteCarlo} propose a dataflow overlay focusing on multi-asset option pricing on AMD Versal platforms.
FEATHER \cite{tong2024feather} mainly focuses on-chip data layout reordering.
NSFlow \cite{yang2025nsflow} is designed for neuro-symbolic AI applications.
In static-dataflow accelerators, CHARM \cite{zhuang2023charm} focuses on the diverse mapping between kernels and accelerators, and CHARM 2.0 \cite{charm2} extends to more data types. SSR \cite{ssr} and EQ-ViT \cite{dong2024eq} are tailored for small models that fit on-chip. 
HMix \cite{thallikar2026hmix} focuses on quantized DNN inference.
DWN \cite{DWN} proposes a hardware-software co-design for edge-compatible high-throughput neural networks.
Multiple deep learning frameworks are supported by hls4ml \cite{hls4ml}, which transforms into high-level synthesis code.
H2PIPE \cite{doumet2024h2pipe} is a dataflow accelerator that can leverage HBM and on-chip storage for layer-pipelined dataflow acceleration.
PEGASO \cite{PEGASO} proposes a specific data layout and AIE mapping for stencil applications, differing from GEMM kernels. 
Menzel et al. \cite{Menzel} propose a dedicated accelerator for electron repulsion integrals, leveraging the specific shuffling features for the first version of AIE.

AI Engine (AIE) is a programmable array of SIMD vector processors integrated in AMD Versal devices to accelerate high-throughput compute workloads, and some works are targeting general matrix multiplication (GEMM) optimization on it. 
In autoMM \cite{dac_automm}, they propose an efficient programming style for matrix multiplication.
MaxEVA \cite{MaxEVA} and GAMA \cite{mhatre2025gama} propose the advanced mapping strategies for AIE and AIE-ML to maximize MM kernel throughput.
Mhatre et al. \cite{perfAnalysis_arora} conduct an analysis targeting the Versal platform based on specialized architecture design for different GEMM kernels.
AIE4ML \cite{danopoulos2025aie4ml} mainly targets small models that can fit AIE tiles. 
However, they are all static dataflow architecture designs.

Other literature focuses on providing programming models to increase the abstraction level of the FPGA, AIE, and heterogeneous accelerator design, including ARIES~\cite{aries}, Allo \cite{chen2024allo}, MLIR-AIR \cite{MLIR-AIR}, and IRON \cite{hunhoff2025iron}.
Chext \cite{sonmez2026chext} proposes a domain-specific language to provide an abstraction layer for complex dataflow circuits.
SPADES \cite{Nguyen2023spades} improves productivity by proposing a design flow for Versal programmable logic.
HiLFS \cite{na2026hilfs} is an HLS-based file system and storage stack that implements storage management entirely on the FPGA.
Xu et al. \cite{Ferroelectric} propose a context-switching FPGA with lower area and power consumption and faster switching latency.
Josipovic et al. \cite{Synthesizing} introduce HLS in a dynamically scheduled form to produce dataflow circuits from imperative code.
In commercial products, AMD Vitis DPU \cite{amd_vitis_ai_dpu} are pre-built closed-source architectures that split on-chip resources into independent homogeneous accelerators, while users cannot explicitly configure the flexible datapath.
Microsoft Brainwave NPU \cite{Brainwave} proposes an NPU architecture targeting batch-1 large RNN models.
Altera~\cite{boutros2020beyond_altera} features tensor blocks for AI-intensive kernels, requiring hardware expertise from programmers.

\vspace{-10pt}
\section{DORA Architecture}
\label{sec: architecture}
In this section, we first introduce the DORA hardware architecture overview in Section \ref{sec: arch overview}, including the components and orchestration for the control plane and data plane. Then, we discuss the management of on-chip storage, computation resources, and the synchronization mechanism in Section~\ref{sec: CU,sec: FMU,sec: IOM}, respectively. Then, we introduce the special function unit for the non-linear kernels in Section \ref{sec: SFU}, and instruction dispatch unit in Section~\ref{sec: IG}. Finally, we present how to generate the DORA architecture in Section~\ref{sec:Architecture Generation}.

\begin{figure}[tbh]
    \centering
    \includegraphics[width=0.7\linewidth]{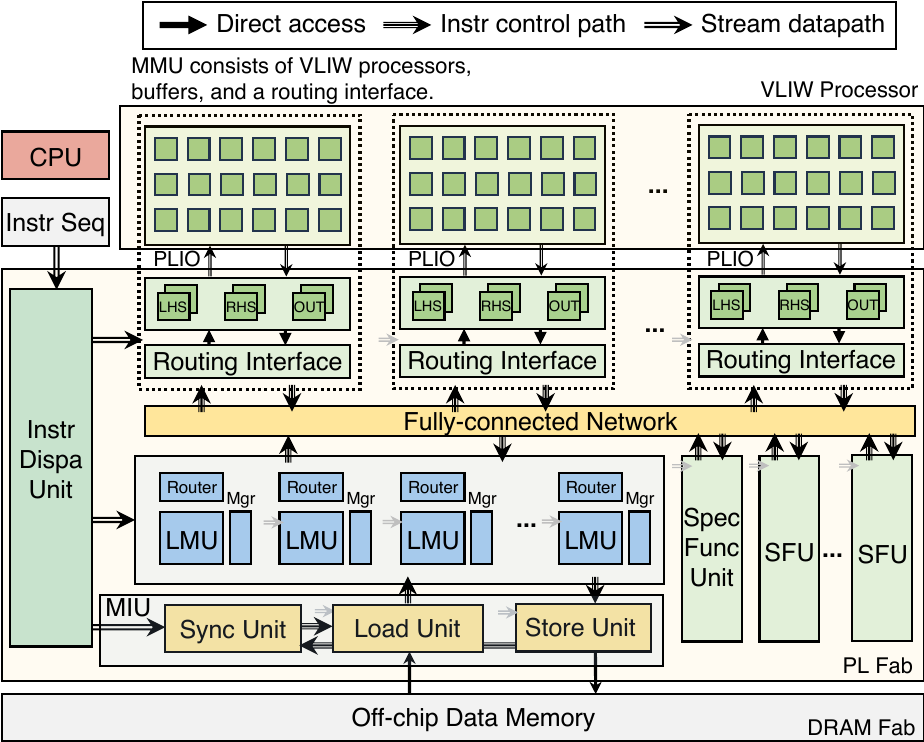}
    \vspace{-10pt}
    \caption{DORA architecture overview.}
    \vspace{-10pt}
    \label{fig: arc overview}
\end{figure}

\vspace{-5pt}
\subsection{Overview}
\label{sec: arch overview}

Figure \ref{fig: arc overview} illustrates the overall DORA hardware architecture.
DORA is a heterogeneous architecture comprising a very long instruction word (VLIW) supported vector processor array to enable efficient GEMM, and traditional programmable logic (PL) to support rapidly evolving non-linear functions and application-specific control-flow customizations.
In the control plane, the host CPU constructs the instruction sequences for each on-chip function unit and loads them into the instruction memory. In the data plane, DORA adopts a fully connected streaming network between units, where instructions can explicitly specify the datapath routing at a fine-grained layer level.

\begin{figure}
    \centering
    \includegraphics[width=0.7\linewidth]{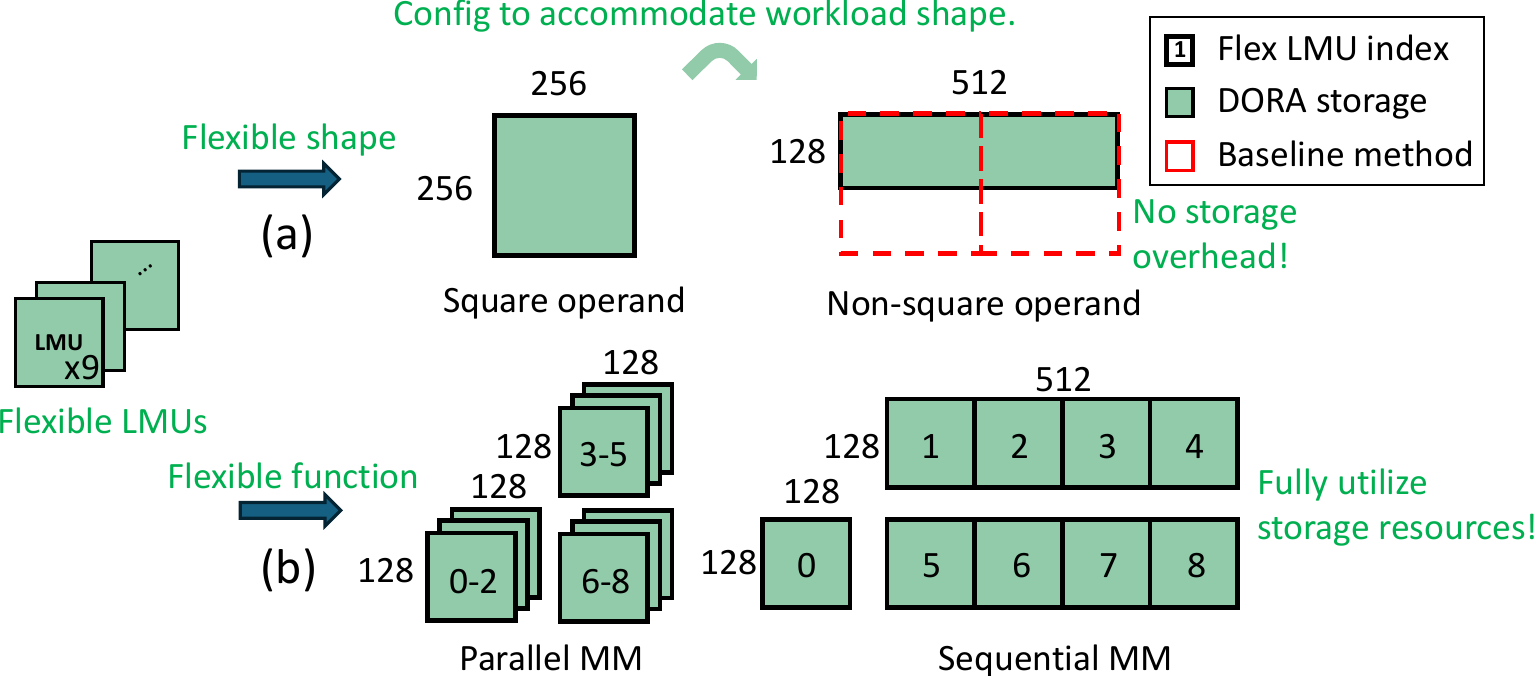}
    \vspace{-8pt}
    \caption{Flexible on-chip memory resource management.}
    \label{fig:FlexLMU}
\end{figure}

\begin{figure}[tbh]
    \centering
    \includegraphics[width=0.8\linewidth]{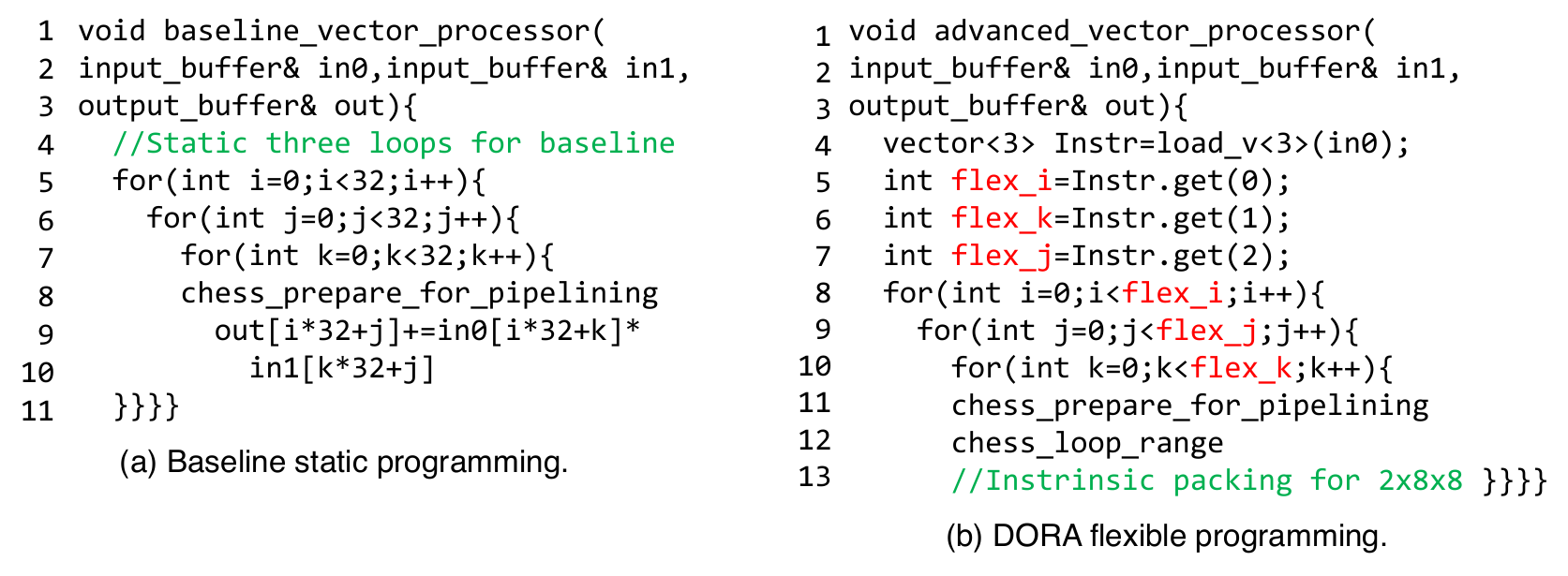}
    \vspace{-10pt}
    \caption{Flexible parallelism management.}
    \vspace{-5pt}
    \label{fig: flex_comp}
\end{figure}

\vspace{-5pt}
\subsection{On-chip Memory Management}
\label{sec: FMU}

Existing accelerators \cite{zhuang2023charm,ssr,chen2024understanding,hall2020hpipe,liu2025flightvgm,zeng2024flightllm,dong2024eq,zhang2020dnnexplorer,kwon2021herald,cai2023set,AMA,dong2024eq} typically pre-allocate dedicated N-dimensional (N-D) on-chip buffers whose dimensions are statically tailored to certain operands and maintain a static dataflow with predefined control, but incur significant overhead on other DNN models.
Assuming there are two different operands with different shapes within one DNN model, i.e., square operand (256$\times$256) and non-square operand (128$\times$512). If the static pre-allocation design methodology is adopted, two separate on-chip buffers need to be allocated. To reduce the required size of on-chip buffers, the on-chip buffer can be tailored to a single operand and reused for the other one at the cost of extra padding, as shown in Figure \ref{fig:FlexLMU}(a). DORA manages on-chip memory at a fine granularity using flexible LMUs, which can be dynamically composed to larger logical buffers, or configured with different functionalities, e.g., LHS, RHS, or OUT, for target operands.

\vspace{-10pt}
\subsection{Computation Parallelism Management}
\label{sec: CU}

DORA can implement matrix multiplication units (MMU) on PL fabric or VLIW-supported vector processor array for higher performance. For PL-based MMU, existing works \cite{chen2024understanding,liu2025flightvgm,zeng2024flightllm,zhang2020dnnexplorer} enable resource optimization for better performance and flexibility.
However, balancing the efficiency in parallelism and flexibility is non-trivial for vector processor-based MMU due to limited intrinsics. Existing approaches typically fall into two categories.
First, some works~\cite{zhuang2023charm,ssr,dong2024eq,MaxEVA,AMA,yang2023aim} program the VLIW processors with static intrinsics and pad input operands to fixed shapes to accommodate diverse workloads (Figure \ref{fig: flex_comp}(a)). Although this approach simplifies control and intrinsic programming, it leads to unnecessary waste in computation.
Second, other works~\cite{RSN} store multiple sets of intrinsic programs in the processors' instruction memory, each tailored to a specific workload, incurring significant memory overhead. 

To address this challenge, DORA enables a dynamic loop boundary methodology that is supported by the VLIW architecture, and combines with optimized intrinsic packing to exploit the instruction-level parallelism, as shown in Figure \ref{fig: flex_comp} (b).
In DORA architecture, each MM can be assigned to multiple MMUs. DORA can adjust the loop boundaries to use different numbers of MMUs for the same MM under different requirements, achieving runtime configurable parallelism management accordingly.


\vspace{-6pt}
\subsection{Off-chip Memory Access Management}
\label{sec: IOM}

During DNN inference, DORA stores intermediate results in off-chip memory (i.e., DRAM) and reloads them when required. This approach accommodates large activation sizes, reduces on-chip buffer requirements, and supports diverse data layouts across different layers. However, this may introduce a read-after-write (RAW) data hazard when a dependent layer attempts to read an intermediate data before the corresponding write to DRAM has completed.

\begin{figure}
    \centering
    \includegraphics[width=0.7\linewidth]{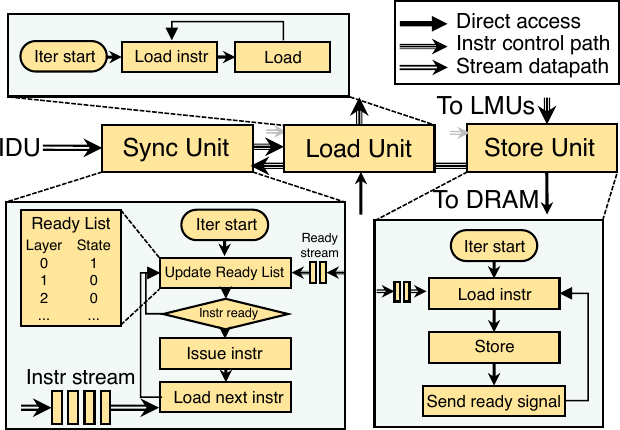}
    \vspace{-10pt}
    \caption{Synchronization mechanism in MIU.}
    
    \label{fig:SyncUnit}
\end{figure}


To address this RAW data hazard, DORA adopts a synchronization mechanism, as illustrated in Figure \ref{fig:SyncUnit}. During the PL kernel execution, the Sync Unit loads instructions from IDU via an instruction stream and maintains a Ready List Table to track each layer's status.
In each iteration, the Sync Unit first checks the ready stream from the Store Unit. Once a layer has been written back to DRAM, the Store Unit emits a ready signal to notify the Sync Unit that the write-back process for a certain layer has completed, and the Sync Unit updates the Ready List Table accordingly. 
Before issuing the read instruction to the Load Unit, the Sync Unit checks whether the dependencies of the current instruction have been resolved. 
If not, it continuously monitors the ready stream until all dependencies are satisfied. Once resolved, the Sync Unit dispatches the instruction to the Load Unit and fetches the next instruction to continue. 


\vspace{-10pt}
\subsection{Special Function Unit}
\label{sec: SFU}

DORA supports various non-linear operators, such as Softmax, GeLU, and LayerNorm via Special Function Unit (SFU) implemented on the PL.
As shown in Figure \ref{fig: arc overview}, to maintain high flexibility when executing non-linear layers, the SFUs are connected with LMUs through the fully-connected network. Based on the execution patterns of non-linear layers, the reduction is performed along the row dimension. Therefore, DORA buffers one row of the matrix and applies non-linear operations in a row-wise manner. 
Once receiving the corresponding instructions, the SFU first loads a tiled row from an LMU and then aggregates multiple tiled rows from multiple LMUs to reconstruct a full matrix row in the line buffer. It then performs the required non-linear operations and streams the results back to the LMUs.
Since this operation is conducted at the tile granularity, the non-linear layers can be pipelined with the linear layers in a tiled fashion to overlap the data communication overhead. 
Moreover, for the super-large layer, where the intermediate data cannot be buffered even in a tiled fashion, DORA treats the non-linear kernel as an independent layer, stores the matrix to off-chip memory, and processes it in a streaming fashion.

\vspace{-10pt}
\subsection{Instruction Dispatch Unit}
\label{sec: IG}


The DORA framework (Section~\ref{sec: compilation framework}) generates the instruction sequence and stores it in the off-chip instruction memory sequentially. As elaborated in Table~\ref{tab: instruction set}, there are two fields in DORA instructions, i.e., instruction headers and instruction bodies.
The instruction header is fixed-width (32 bits in DORA), while the instruction body is variable-width and unit-specific, as different function units require different metadata.
Instruction headers identify the type of the subsequent instruction body and encode metadata such as \texttt{des\_unit} indicating the destination unit for the current instruction, shown as IDU in Table \ref{tab: instruction set}. Instruction bodies, shown as MIU, SFU, LMU, and MMU, specify the unit-specific control information required to orchestrate flexible dataflow.
During the runtime, IDU first fetches instruction headers from instruction memory, decodes the instruction header, loads several following bytes according to \texttt{valid\_length}, and dispatches them to function units based on \texttt{des\_unit}.
Each function unit continuously loads and decodes instructions until receiving \texttt{is\_last} signal, and decodes the required information in instruction items for dataflow orchestration. For example, MMU instruction carries \texttt{ping\_op} and \texttt{pong\_op} to decide the operating buffers, \texttt{bound\_i}, \texttt{bound\_k}, and \texttt{bound\_j} to achieve flexible parallelism for VLIW processors, \texttt{src\_lmu} and \texttt{des\_lmu} to route the datapath.

\begin{table}[t]
\caption{DORA instruction set.}
\vspace{-10pt}
\centering
\scriptsize
\setlength{\tabcolsep}{3pt}
\renewcommand{\arraystretch}{1.05}
\resizebox{\columnwidth}{!}{
\begin{tabular}{l l p{2.8cm} l p{2.8cm}}
\toprule
\textbf{Unit} & \textbf{Field} & \textbf{Description} & \textbf{Field} & \textbf{Description} \\
\midrule

\multirow{2}{*}{\textbf{IDU}}
& \texttt{is\_last}      & Last instruction flag
& \texttt{op\_type}      & Operation type \\
& \texttt{des\_unit}     & Destination unit
& \texttt{valid\_length} & Valid body length \\
\midrule

\multirow{3}{*}{\textbf{MIU}}
& \texttt{ddr\_addr}     & Off-chip memory address
& \texttt{src\_lmu}      & Source LMU index \\
& \texttt{des\_lmu}      & Destination LMU index
& \texttt{M}, \texttt{N} & Tile dimensions \\
& \texttt{start\_row}, \texttt{end\_row} & Row range
& \texttt{start\_col}, \texttt{end\_col} & Column range \\
\midrule

\multirow{2}{*}{\textbf{SFU}}
& \texttt{src\_lmu}      & Source LMU index
& \texttt{des\_lmu}      & Destination LMU index \\
& \texttt{count}         & Operation count
& \texttt{ele\_num}      & Element count \\
\midrule

\multirow{3}{*}{\textbf{LMU}}
& \texttt{ping\_buf}, \texttt{pong\_buf} & Ping-pong buffer selection
& \texttt{load\_op}, \texttt{send\_op}   & Load/send control \\
& \texttt{src\_pu}, \texttt{des\_pu}     & Source/destination PU
& \texttt{count}                         & Transfer count \\
& \texttt{start\_row}, \texttt{end\_row} & Row range
& \texttt{start\_col}, \texttt{end\_col} & Column range \\
\midrule

\multirow{2}{*}{\textbf{MMU}}
& \texttt{ping\_op}, \texttt{pong\_op}   & Ping-pong op control
& \texttt{bound\_i}, \texttt{bound\_k}, \texttt{bound\_j} & Loop bounds \\
& \texttt{src\_lmu}      & Source LMU index
& \texttt{des\_lmu}      & Destination LMU index \\
\bottomrule
\end{tabular}
}
\label{tab: instruction set}
\vspace{-6pt}
\end{table}

\vspace{-8pt}
\subsection{Architecture Generation}
\label{sec:Architecture Generation}
DORA enables rapid hardware customization while preserving a unified programming and execution model. DORA is designed in a modular fashion, and users can generate customized DORA architecture using a template-based design approach, where users only need to specify the number of different function units, such as MMU, LMU, etc., according to the application requirements and resource constraints. 

To accommodate the rapid evolution of non-linear functions in modern DNN models, DORA supports the integration of user-defined non-linear functions developed in HLS C/C++, which are then synthesized and instantiated as SFUs, allowing them to be seamlessly incorporated into the DORA execution pipeline. 
By combining the template-based architecture generation with HLS-based extensibility, DORA enables both architectural scalability and functional adaptability, making it suitable for a wide range of DNN workloads and deployment scenarios.

\vspace{-5pt}
\section{DORA compilation framework}
\label{sec: compilation framework}

In this section, we first present the DORA compilation framework and workflow in Section \ref{sec:framework overview}. Then, we explain the two-stage design space exploration (DSE), including performance modeling in Section \ref{sec:perf model} and formulations in Sections \ref{sec:MILP} and \ref{sec:GA}.

\vspace{-5pt}
\subsection{Framework Overview}
\label{sec:framework overview}

\begin{figure}
    \centering
    \includegraphics[width=0.7\linewidth]{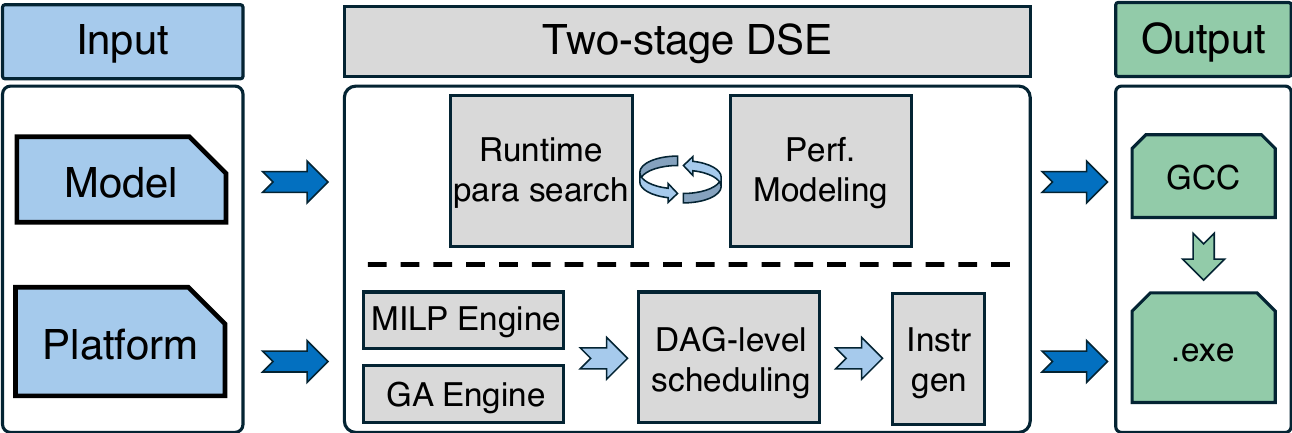}
    \vspace{-10pt}
    \caption{DORA framework overview.}
    \vspace{-10pt}
    \label{fig: framework}
\end{figure}

Figure \ref{fig: framework} illustrates the DORA compilation framework, which consists of a two-stage DSE optimizer and an automatic instruction generator. DORA takes the target DNN model, DRAM profiling results, and platform specifications as inputs. It first performs a two-stage DSE to generate an execution timeline for the workload, followed by instruction generation for deployment.
In the first-stage DSE, DORA applies a performance model (Section \ref{sec:perf model}) to estimate per-layer latency under different runtime parameters, including off-chip bandwidth, tile sizes, and computation parallelism. During this process, DORA records optimal configurations under different resource constraints for each layer and constructs a candidate execution table.
In the second-stage DSE, DORA formulates a MILP model based on \cite{milp_basic, zhou2019dissertation} to explore parallel execution opportunities. Based on the resulting execution timeline, DORA generates unit-wise instruction sequences and invokes the backend compiler to produce executable files.

\vspace{-8pt}
\subsection{Performance Modeling}
\label{sec:perf model}

In the first-stage DSE, given a matrix multiplication (MM) of size $M \times K \times N$ and resource constraints on \texttt{\#ReqLMU}, \texttt{\#ReqMMU}, and \texttt{\#ReqSFU}, DORA explores the runtime parameter space and selects the optimal configuration to populate the candidate execution table.

Based on the discussion in Section~\ref{sec: architecture}, DORA supports multiple levels of flexible tile sizes by adjusting computation parallelism. Specifically, each vector processor computes a tiled MM of \texttt{aie\_m} $\times$ \texttt{aie\_k} $\times$ \texttt{aie\_n}. Within one MMU, the vector processor array follows a $4 \times 4 \times 4$ composition pattern, which is searched from DSE and must be fixed at compile time due to static PLIO routing. Besides, multiple MMUs can also be aggregated into a large MMU, along the non-reduced dimension, with the composing pattern of \texttt{MMU\_m} $\times$ 1 $\times$ \texttt{MMU\_n}. Therefore, the computation tile size for one processor launch iteration is \texttt{(aie\_m} $\times$ $4$ $\times$ \texttt{MMU\_m)} $\times$ \texttt{(aie\_k} $\times$ $4$ $\times$ $1$\texttt{)} $\times$  \texttt{(aie\_n} $\times$ $4$ $\times$ \texttt{MMU\_n)}.

In terms of on-chip storage mapping, we adopt the minimum mapping granularity of one MMU tile, i.e.,  \texttt{(aie\_m} $\times$ $4$ \texttt{)} $\times$ \texttt{(aie\_k} $\times$ $4$\texttt{)} $\times$  \texttt{(aie\_n} $\times$ $4$\texttt{)}. 
By enumerating feasible MMU tile configurations, we determine the on-chip tile size of \texttt{LMU\_m} $\times$ \texttt{LMU\_k} $\times$ \texttt{LMU\_n}.
Moreover, since each LMU supports flexible tile shapes, we further determine the operand-specific storage patterns.
Specifically, for the LHS operand, we determine the number of MMU tiles buffered in each LMU along the row and column dimensions, denoted as \texttt{lhs\_MMU\_row} $\times$ \texttt{lhs\_MMU\_col}. Similarly, the storage parameters for the RHS and OUT operands can be derived accordingly.
Therefore, given the off-chip bandwidth, the stream port width in the fully connected network, and the tile sizes of MMUs and LMUs, we can derive the latency of a single vector processor iteration, denoted as \texttt{latency\_MMU}. Accordingly, we compute the latency of one on-chip data-reuse iteration \texttt{latency\_LMU}, and the total latency is given by \texttt{latency\_LMU} $\times$ \texttt{iter\_times}, where \texttt{iter\_times} = \texttt{(M/LMU\_m)*(K/LMU\_k)*(N/LMU\_n)}.

\vspace{-10pt}
\subsection{MILP Formulation}
\label{sec:MILP}

\begin{figure}
    \centering
    \includegraphics[width=1\linewidth]{algo.pdf}
    \vspace{-20pt}
    \caption{MILP formulation.}
    \label{fig: algo}
\end{figure}

In Section \ref{sec:perf model}, Stage 1 ensures optimality by exhaustively enumerating all possible layer-to-accelerator mapping candidates. Based on the candidate table, Stage 2 is formulated as an MILP problem to optimize scheduling.

As shown in Figure \ref{fig: algo}, at the workload graph level, one node corresponds to a layer ($L_i$), and one edge corresponds to an inter-layer dependency ($P_{i,j}=1$). $l_{i,k}$, $m_{i,k}$, and $s_{i,k}$ represent the required number of LMU, MMU, and SFU, respectively, if $L_i$ executed under mode $M_{i,k}$ with the optimal latency $e_{i,k}$.The objective is to derive an optimal scheduling timeline that minimizes the latency while satisfying both dependency and resource constraints (Lines 2 and 15).
$A_{i,m}$, $B_{i,m}$, and $C_{i,m}$ are binary variables, to represent whether $L_i$ is assigned the m-th LMU, MMU, or SFU. $L_i$ starts at $S_i$ and ends at $E_i$.
$L_i$ only has one feasible execution mode (Line 4). 
There is a precedence constraint on the layers with $P_{i,j}=1$ (Line 5), and $E_i$ is calculated by Line 6.
Besides, each functional unit can process only one layer at a time, and arbitrary two layers on the same unit must not overlap temporally, which can be modeled by an overlap variable $O_{i,j}$ ($1$, if $S_i - E_j < 0$). Then, we linearize it into inequalities (Lines 7$\sim$8) by introducing a sufficiently large constant $\phi$.
Therefore, $O_{i,j}=O_{j,i}=1$ represents two layers overlap, and for any pair of independent layers ($P_{i,j}=0$) mapped to the same unit, Lines 9$\sim$11 hold. Lines 12$\sim$14 guarantee the resource requirements.

\vspace{-10pt}
\subsection{DSE Acceleration Options}
\label{sec:GA}

\noindent\textbf{DAG Partition.} DORA framework supports two DSE strategies. One option is to partition the workload DAG into several sub-DAGs and launch DSE engines in parallel to optimize each sub-DAG independently (partitioned DSE). Alternatively, the entire DAG can be optimized as a whole (non-partitioned DSE). Theoretically, non-partitioned DSE has the opportunity for global optimization across all the layers. However, under a limited search time budget, the preferred strategy may vary in practice, as evaluated in Section \ref{sec: exp_dse}.

\noindent\textbf{Genetic Algorithm.} The main challenge stems from the extremely large design space induced by the large DAG size and the number of candidates, since the complexity is $O(m^n)$ where $m$ and $n$ equal candidate mode numbers and layer numbers.
To address the scalability issue, we propose a heuristic-based genetic algorithm for the Schedule Optimizer in the DORA framework that encodes decision variables into a chromosome and initializes the population with a predefined size. Through crossover and mutation, new chromosomes are generated and evaluated using a fitness function, and the best individuals are selected for the next generation. In DORA, each design point is represented by a chromosome with $2N$ variables, where $N$ is the number of layers in the DAG. The first $N$ variables, denoted as \texttt{Encode[N]}, are real numbers in $[0,1]$ representing scheduling priorities, while the remaining $N$ variables, \texttt{Candidate[N]}, are integers indicating the selected execution modes. During decoding, a dependency-aware decoder enforces precedence constraints to generate feasible schedules from the encoded priorities and execution modes.

\vspace{-10pt}
\section{A DORA Case Study}
\label{sec: case study}

In this section, we introduce the DORA compilation framework workflow, including the two-stage DSE optimization and instruction compilation process in Section \ref{sec: casestudy DSE compilation}. 
We then introduce the DORA architecture control flow and data flow at runtime in Section \ref{sec: casestudy runtime}.

\vspace{-10pt}
\subsection{DSE and Instruction Compilation Workflow}
\label{sec: casestudy DSE compilation}

\begin{figure}
    \centering
    \includegraphics[width=0.7\linewidth]{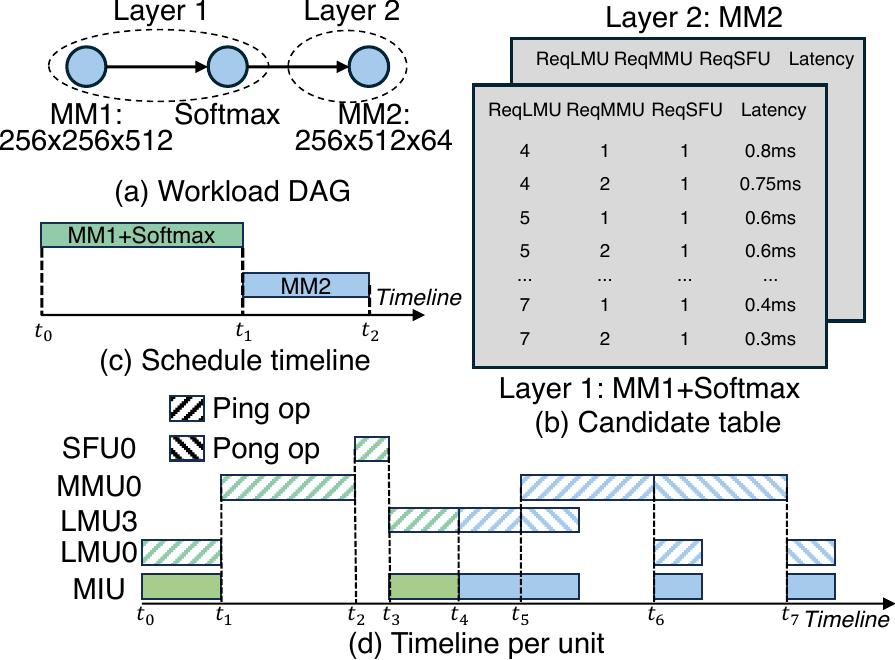}
    \vspace{-10pt}
    \caption{DORA compilation and runtime behaviors.}
    \vspace{-5pt}
    \label{fig:case study}
\end{figure}

As shown in Figure~\ref{fig:case study}(a), we consider a DNN workload with three kernels, where Softmax depends on MM1 and MM2 depends on Softmax. We assume the target hardware contains $7$ LMUs, $2$ MMUs, and $1$ SFU. In the first-stage DSE, the workload is segmented into layers, each categorized as either an MM-only kernel or an MM kernel followed by a non-linear kernel. The performance model estimates latency under different resource budgets.
For Layer 1 (MM1 + Softmax), at least $4$ LMUs are required to buffer the LHS, RHS, OUT operands, and the non-linear kernel output. At least $1$ MMU is required since the MM can be executed in a tiled manner. Consequently, Layer 1 produces $8$ candidate resource configurations, shown in Figure~\ref{fig:case study}(b). The candidate table and platform resource constraints are then passed to the second-stage DSE to explore parallelism.
The two-stage optimizer constructs a feasible schedule satisfying data dependencies, producing an optimal (MILP) or near-optimal (GA) makespan in Figure~\ref{fig:case study}(c). The scheduling timeline encodes both execution order and selected runtime parameters from the candidate table, including the mapping of operand tiles to function units. Based on the layer-wise schedule in Figure~\ref{fig:case study}(c), we derive the function-unit timeline in Figure~\ref{fig:case study}(d), which highlights the behaviors of MIU, LMU0, LMU3, MMU0, and SFU0.

\vspace{-10pt}
\subsection{Runtime Dataflow and Control Flow}
\label{sec: casestudy runtime}

The instructions generated in Section~\ref{sec: casestudy DSE compilation} are pre-stored in DORA’s off-chip memory. At runtime, the IDU fetches instructions through the instruction DRAM port and dispatches them starting at $t_0$. Meanwhile, the MIU and LMU0 receive their instructions and execute them in a synchronized manner. The MIU first loads a $256\times256$ tile (the LHS operand of MM1) from DRAM and streams it to LMU0.
After executing the first instruction, LMU0 immediately fetches the next instruction that forwards the tile to MMU0 at $t_1$. Before $t_1$, MMU0 has already received a read instruction on the LMU0$\rightarrow$MMU0 stream. Since the stream is empty before $t_1$, MMU0 stalls due to back-pressure. This stream-based synchronization mechanism naturally enforces correct on-chip dataflow timing and ordering.
Moreover, Layer 2 has a RAW dependency on Layer 1. As the load instruction of Layer 2 is placed immediately after the store instruction of Layer 1 in the MIU instruction stream, the synchronization mechanism delays issuing the load until the write-back of Layer 1 completes at $t_4$.

\vspace{-10pt}
\section{Experiments}

We prototype the DORA architecture on the AMD Versal VCK190~\cite{versal_acap} platform with 150MHz on PL and 1GHz on AI Engine (AIE). 
We search the hyperparameters, including the AIE composing pattern and the number of instantiated functional units, and instantiate $6$ MMUs each with $4$$\times$$4$$\times$$4$ AIE tiles, $14$ LMUs, and $3$ SFUs in the DORA architecture.
To be noticed, they are not for a specific model but for a series of model sets.
We first present the PL resource utilization for DORA architecture in Section~\ref{sec: exp_resource}
Then, we evaluate the effectiveness of the proposed dynamic computation parallelism management methodology for a single VLIW processor using AIE in Section~\ref{sec: exp_singleAIE}. 
We evaluate the end-to-end performance on diverse workloads in Section~\ref{sec: exp_end2end} by comparing with CHARM 2.0~\cite{zhuang2023charm} and RSN~\cite{RSN}, which are state-of-the-art architectures in fixed data path accelerator and overlay accelerator, respectively. Among AutoMM~\cite{autoMM}, CHARM \cite{zhuang2023charm}, CHARM 2.0 \cite{charm2}, and ARIES \cite{aries}, we adopt the open-source implementation of CHARM 2.0 and reproduce the two-diverse designs for given models, which achieves the best baseline performance. For RSN, since they don’t open-source analytical models and usage methods, we build an in-house analytical model that follows proposed features.
We also compare different DSE strategies and demonstrate the effectiveness of DORA DSE engines in Section~\ref{sec: exp_dse}.

\vspace{-10pt}
\subsection{Hardware Resource Utilization}
\label{sec: exp_resource}

Figure~\ref{fig:layout} depicts the prototyped DORA hardware layout on Versal VCK190 platform. We omit to highlight fully-connected network since it spans all PL fabric on FPGA.
As summarized in Table~\ref{tab: exp_resource_breakdown}, the DORA architecture effectively utilizes the available AIEs and URAM resources, while the additional control logic only consumes redundant resources such as REGs and LUTs.
These results demonstrate that DORA is readily deployable on existing reconfigurable platforms effectively.

\vspace{-10pt}
\subsection{Efficiency on Single AIE Flexible Parallelism}
\label{sec: exp_singleAIE}

\begin{figure}[t]
    \centering

    \begin{minipage}[t]{0.3\linewidth}
        \centering
        \includegraphics[width=\linewidth]{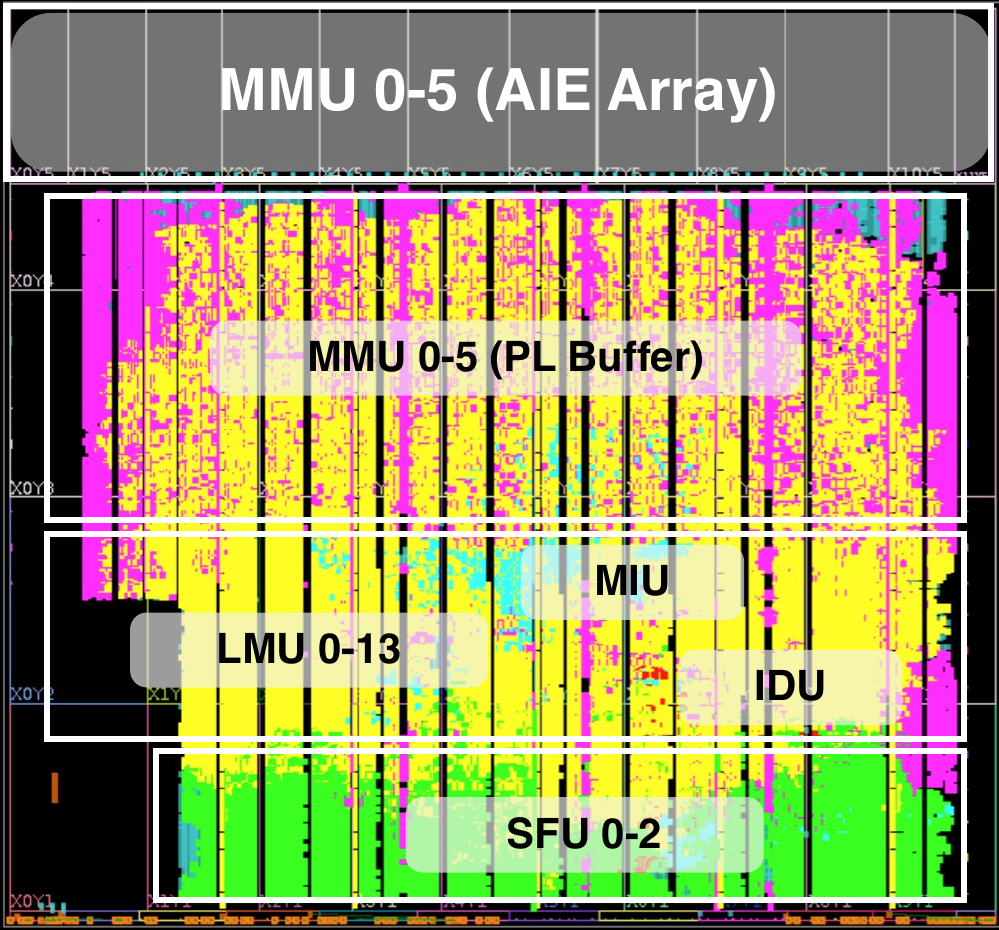}
        \vspace{-20pt}
        \caption{Device layout of DORA architecture.}
        \label{fig:layout}
    \end{minipage}
    \hfill
    \begin{minipage}[t]{0.68\linewidth}
        \centering
        \includegraphics[width=\linewidth]{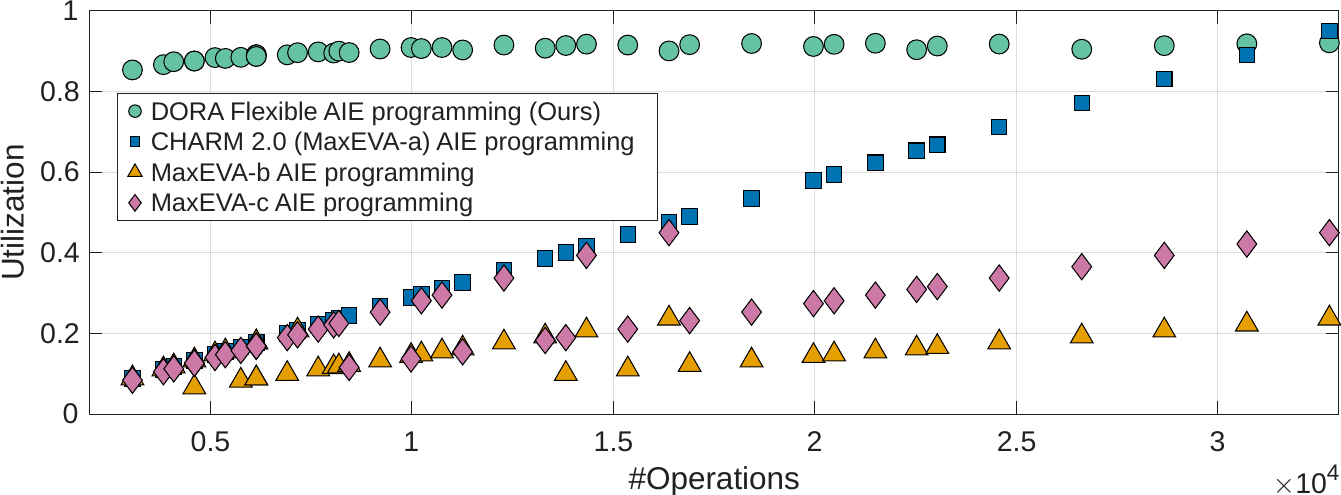}
        \vspace{-20pt}
        \caption{Single AIE efficiency under \#operations variation.}
        \label{fig:exp_singleAIE}
    \end{minipage}
\end{figure}

We demonstrate the single AIE computational efficiency gains enabled by our proposed flexible parallelism management, using an FP32 MM microbenchmark with varying problem sizes. We implement the single AIE kernel using AIE intrinsics \cite{amd_aie_intrinsics_ug} and measure the cycles on the Versal ACAP AIE System C simulator \cite{amd_aie_systemc_sim}. 
We compare DORA with CHARM 2.0 \cite{charm2} and MaxEVA \cite{MaxEVA}, and sweep the MM dimensions from $8\times24\times16$ to $32\times32\times 32$. CHARM 2.0 keeps each AIE tile to $32\times32\times32$, while MaxEVA provides multiple AIE tile sizes, and we select $32\times32\times32$, $16\times128\times16$, and $16\times32\times64$ as MaxEVA-a, MaxEVA-b, and MaxEVA-c, respectively.
As shown in Figure~\ref{fig:exp_singleAIE}, the results demonstrate that our design sustains a wide range of MM sizes with up to a $6\times$ variation in operation counts, while maintaining efficiency variation within 5\%.
CHARM 2.0 and MaxEVA can only obtain efficient resource utilization on specific MM operation shapes, i.e., \texttt{M}, \texttt{K}, and \texttt{N} are multiples of AIE tile sizes, while incurring much overhead in diverse MM operation shapes.
The proposed dynamic loop boundaries introduce negligible overhead, and DORA incurs only about $1\%$ degradation at Point~\textcircled{b} and consistently outperforms CHARM on all other cases, achieving up to 8$\times$ efficiency improvement.


\vspace{-10pt}
\subsection{Performance on End-to-End DNN Workload}
\label{sec: exp_end2end}

\begin{figure}
    \centering
    \includegraphics[width=0.8\linewidth]{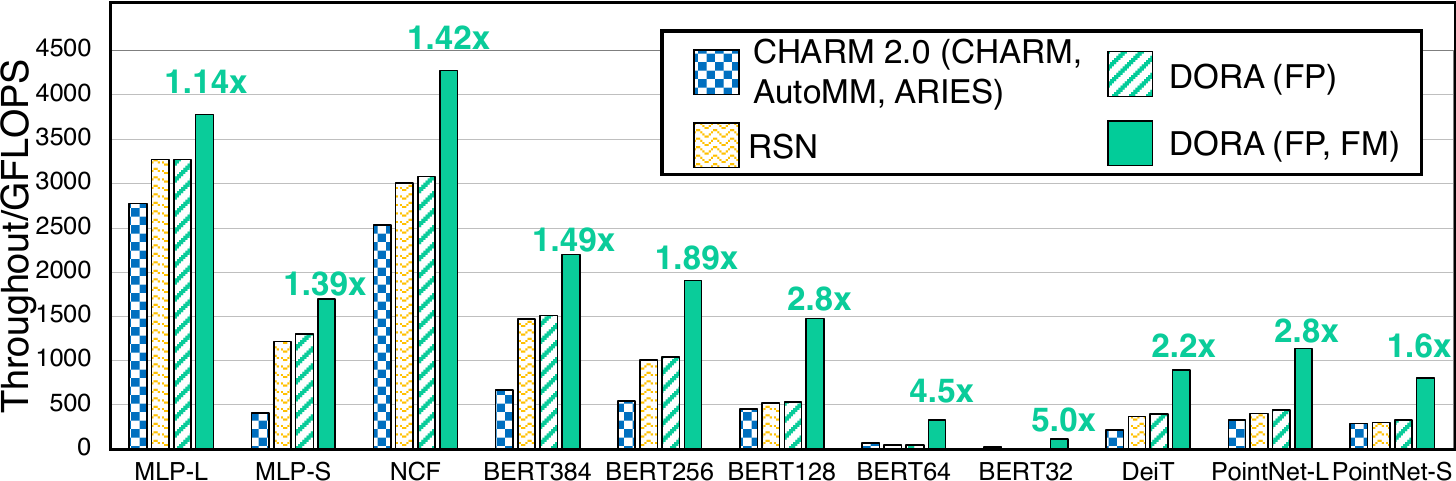}
    \vspace{-12pt}
    \caption{End-to-end performance and gains. FP: flexible parallelism; FM: flexible memory management.}
    \vspace{-5pt}
    \label{fig: end-to-end exp}
    
\end{figure}

To evaluate DORA's adaptability across diverse workloads, we apply DORA to a series of realistic models with varying model sizes from 0.8M to 110M, including MLP \cite{mlp}, DeiT \cite{DeiT}, BERT \cite{Bert}, PointNet \cite{qi2017pointnet}, and NCF \cite{ncf} in FP32 data type.
As shown in Figure~\ref{fig: end-to-end exp}, the x-axis lists different workloads, the left y-axis reports throughput in GFLOPS, and the right y-axis shows DORA’s throughput improvement over RSN~\cite{RSN}. Overall, DORA achieves up to a 5$\times$ throughput improvement compared to the best-performing case of CHARM or RSN. Moreover, to evaluate the effectiveness of DORA’s on-chip storage and computation-parallelism management, we perform an ablation study on three mechanisms, i.e., flexible MMU parallelism and flexible LMU memory functionalities.
DORA consistently outperforms CHARM 2.0 and RSN across all model sizes, with particularly large gains on small models. MLP consists of large, near-square MM operations ($3072\times4096\times4096$), which can saturate on-chip memory and parallelism, resulting in computation-bound and small gains. However, NCF consists of diverse MM shapes even with $3072\times32\times1$, introducing much imbalance between operands, which can provide optimization opportunities for flexible LMU functionality to achieve the maximum data reuse. BERT-32 is a tiny model with small MM layer shapes, and DORA can configure each LMU to match with operands in a fine-grained fashion, to avoid extra data padding overhead and off-chip bandwidth waste.
Moreover, the benefit of flexible parallelism management is obvious in small models since the tile size saturates the parallelism in large models, and flexible memory management provides substantial performance improvements across all model sizes since it can adapt the flexible dataflow at layer granularity to reduce overhead.
In contrast, CHARM 2.0 and RSN incur inefficiency due to padding overhead and underutilized parallelism for small layers.

\vspace{-10pt}
\subsection{DSE Acceleration Options Evaluation}
\label{sec: exp_dse}
\vspace{-3pt}

\begin{table}[t]
\centering
\caption{DORA resource utilization breakdown.}
\vspace{-8pt}
\label{tab:exp_resource_breakdown}
\resizebox{0.6\linewidth}{!}{%
\setlength{\arrayrulewidth}{0.4pt}%
\begin{tabular}{|c|c|c|c|c|c|c|c|}
\hline
Unit &
  REG &
  LUTLogic &
  LUTMem &
  BRAM &
  URAM &
  DSP &
  AIE \\ \hline
MMU &
  \begin{tabular}[c]{@{}c@{}}110598\\ (6.12\%)\end{tabular} &
  \begin{tabular}[c]{@{}c@{}}103890\\ (11.52\%)\end{tabular} &
  0 &
  \begin{tabular}[c]{@{}c@{}}576\\ (60\%)\end{tabular} &
  0 &
  \begin{tabular}[c]{@{}c@{}}6\\ (0.3\%)\end{tabular} &
  \begin{tabular}[c]{@{}c@{}}384\\ (96\%)\end{tabular} \\ \hline
LMU &
  \begin{tabular}[c]{@{}c@{}}64358\\ (3.64\%)\end{tabular} &
  \begin{tabular}[c]{@{}c@{}}174538\\ (19.46\%)\end{tabular} &
  \begin{tabular}[c]{@{}c@{}}43358\\ (9.66\%)\end{tabular} &
  0 &
  \begin{tabular}[c]{@{}c@{}}448\\ (96\%)\end{tabular} &
  \begin{tabular}[c]{@{}c@{}}280\\ (14.28\%)\end{tabular} &
  0 \\ \hline
SFU &
  \begin{tabular}[c]{@{}c@{}}72000\\ (4\%)\end{tabular} &
  \begin{tabular}[c]{@{}c@{}}100974\\ (11.22\%)\end{tabular} &
  \begin{tabular}[c]{@{}c@{}}2328\\ (0.5\%)\end{tabular} &
  \begin{tabular}[c]{@{}c@{}}48\\ (4.95\%)\end{tabular} &
  0 &
  \begin{tabular}[c]{@{}c@{}}96\\ (4.89\%)\end{tabular} &
  0 \\ \hline
IDU &
  \begin{tabular}[c]{@{}c@{}}613\\ (0.03\%)\end{tabular} &
  \begin{tabular}[c]{@{}c@{}}415\\ (0.05\%)\end{tabular} &
  0 &
  0 &
  0 &
  0 &
  0 \\ \hline
MIU &
  \begin{tabular}[c]{@{}c@{}}1810\\ (0.09\%)\end{tabular} &
  \begin{tabular}[c]{@{}c@{}}3913\\ (0.44\%)\end{tabular} &
  \begin{tabular}[c]{@{}c@{}}28\\ (\textless{}0.1\%)\end{tabular} &
  0 &
  0 &
  \begin{tabular}[c]{@{}c@{}}6\\ (0.3\%)\end{tabular} &
  0 \\ \hline
\begin{tabular}[c]{@{}c@{}}Fully-con\\ Network\end{tabular} &
  \begin{tabular}[c]{@{}c@{}}264768\\ (14.7\%)\end{tabular} &
  \begin{tabular}[c]{@{}c@{}}68208\\ (7.58\%)\end{tabular} &
  0 &
  0 &
  0 &
  0 &
  0 \\ \hline
Others &
  \begin{tabular}[c]{@{}c@{}}55111\\ (3\%)\end{tabular} &
  \begin{tabular}[c]{@{}c@{}}65504\\ (7.28\%)\end{tabular} &
  \begin{tabular}[c]{@{}c@{}}1419\\ (0.3\%)\end{tabular} &
  \begin{tabular}[c]{@{}c@{}}23\\ (2.4\%)\end{tabular} &
  0 &
  0 &
  0 \\ \hline
Total &
  \begin{tabular}[c]{@{}c@{}}569258\\ (31.5\%)\end{tabular} &
  \begin{tabular}[c]{@{}c@{}}517442\\ (57.4\%)\end{tabular} &
  \begin{tabular}[c]{@{}c@{}}47133\\ (10.5\%)\end{tabular} &
  \begin{tabular}[c]{@{}c@{}}647\\ (67.4\%)\end{tabular} &
  \begin{tabular}[c]{@{}c@{}}448\\ (96\%)\end{tabular} &
  \begin{tabular}[c]{@{}c@{}}388\\ (19.8\%)\end{tabular} &
  \begin{tabular}[c]{@{}c@{}}384\\ (96\%)\end{tabular} \\ \hline
\end{tabular}%
}
\end{table}


\begin{figure}
    \centering    \includegraphics[width=0.8\linewidth]{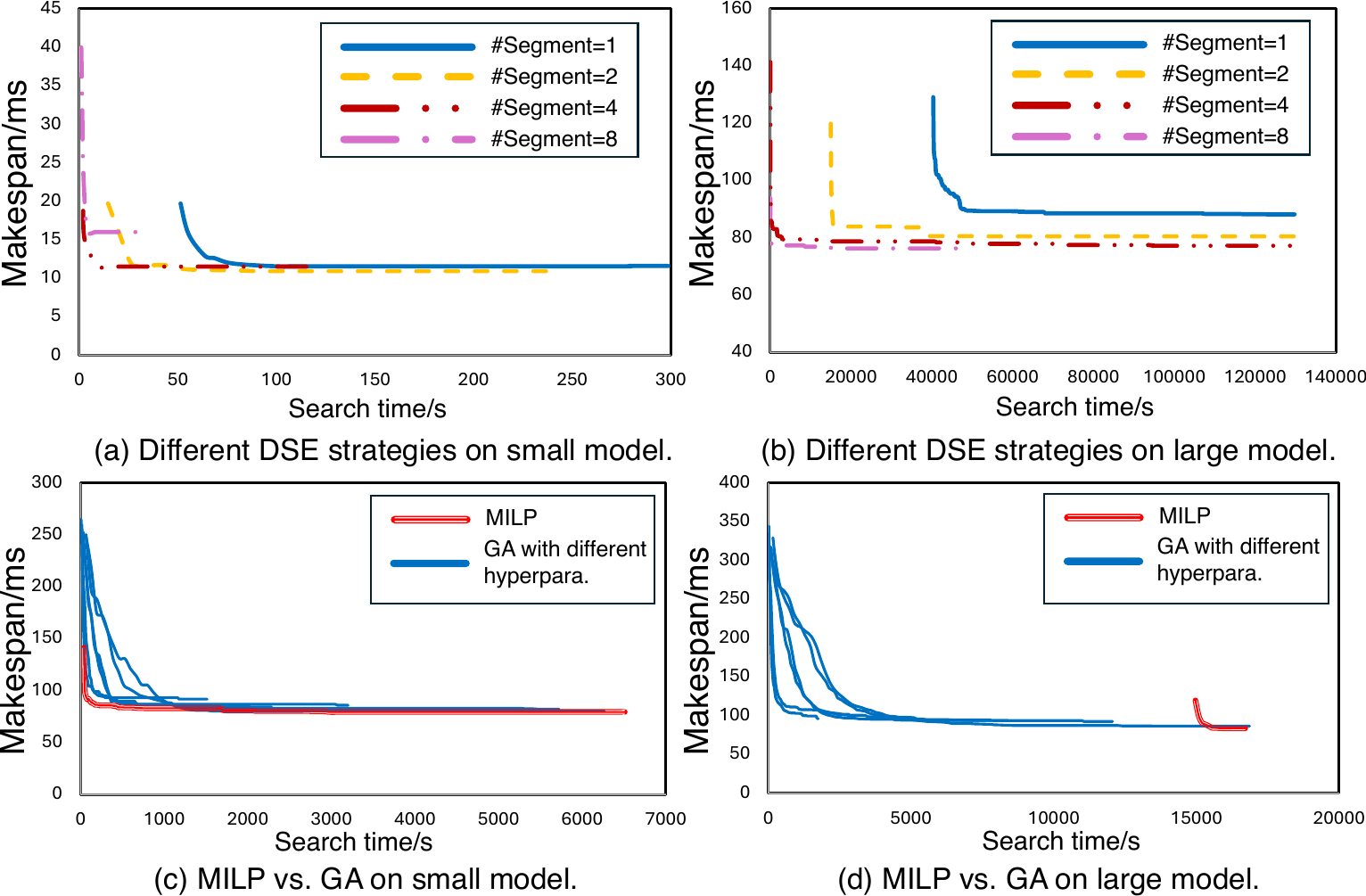}
    \vspace{-10pt}
    \caption{DSE acceleration options evaluation.}
    \vspace{-10pt}
    \label{fig: exp_dse}
\end{figure}








To demonstrate the effectiveness of DSE engines and compare the influence of different DSE strategies, we conduct a series of experiments on an Intel Xeon Gold 6346 CPU server, as shown in Figure \ref{fig: exp_dse}. We apply CPLEX \cite{cplex} and Pymoo \cite{pymoo} as DORA DSE solvers and assign one CPU thread per search process.
In Figure~\ref{fig: exp_dse}(a) and (b), we evaluate DAG partitioning (Section~\ref{sec:GA}) on different DNN models.
For both small and large MLP models with 16 and 128 layers, respectively, the DAG partition option can effectively accelerate the MILP approach. As the number of segments increases, DORA can generate near-optimal scheduling solutions earlier in the search process. In contrast, without DAG partitioning (\#Segment=1), the schedule quality for the large model stagnates in the limited time budget. In practice, users can apply DGA partitioning for large DNN models based on their hardware capacity and assign each segment to a single CPU thread.
Moreover, Figures \ref{fig: exp_dse}(c) and (d) compare a series of GA searches with different hyperparameters (Section~\ref{sec:GA}), against the MILP-based search. Overall, the MILP engine achieves better performance with shorter search time for small models. However, for large models, the MILP engine may fail to find a feasible solution even with a long time budget. In practice, users can apply the MILP engine for small workloads and switch to the GA engine for large workloads.

\vspace{-6pt}
\section{Generality of DORA}
\label{sec:generality}
\vspace{-5pt}
In terms of supporting models, DORA supports the Transformer-based models with varying sequence length, different operation kernels, including MM, Softmax, LayerNorm, and GeLU kernels, and different functional models. DORA leverages off-chip memory and tile-level pipelining for models with larger parameter sizes and sequence lengths. Besides, in terms of supporting platforms, DORA supports different platform backends, including PL-only and PL+AIE platforms from AMD and Altera, by instantiating MMU on the PL or AIE fabric and other functional units on the PL fabric, then reusing the fully-connected network to maintain the same functionality, and launching DSE for instruction generation.

\vspace{-6pt}
\section{Conclusion}
\vspace{-5pt}
This paper proposes an instruction-based overlay architecture with a compilation framework, enabling fine-grained control of data movement, computation, and synchronization for DNN acceleration, which can achieve up to 5$\times$ performance improvement and 90\% scheduling optimality under practical time constraints.

\smallskip
{\small
{\noindent\textbf{ACKNOWLEDGEMENTS --}} This work is supported in part by Brown University New Faculty Start-up Grant, DOE award DE-SC0026344,
NSF awards 
\#2140346, 
\#2231523, 
\#2441179, 
\#2348306, 
\#2511445, 
\#2518375, 
\#2536952, 
\#2544032. 
We thank AMD for the hardware and software donations.
P. Zhou has a financial interest in Shanghai Suikun.
\vspace{-10pt}

\bibliographystyle{ACMReferenceFormat}
\bibliography{Ref_X_short}

\end{document}